# How S-S' di quark pairs signify an Einstein constant dominated cosmology, and lead to new inflationary cosmology physics.

A. W. Beckwith

**ABSTRACT**

We review the results of a model of how nucleation of a new universe occurs, assuming a di quark identification for soliton-anti soliton constituent parts of a scalar field. Initially, we employ a false vacuum potential system; however, when cosmological expansion is dominated by the Einstein cosmological constant at the end of chaotic inflation, the initial di quark scalar field is not consistent w.r.t a semi classical consistency condition we analyze as the potential changes to the chaotic inflationary potential utilized by Guth . We use Scherrer's derivation of a sound speed being zero during initial inflationary cosmology, and obtain a sound speed approaching unity as the slope of the scalar field moves away from a thin wall approximation. All this is to aid in a data reconstruction problem of how to account for the initial origins of CMB due to dark matter since effective field theories as presently constructed require a cut off value for applicability of their potential structure. This is often at the cost of, especially in early universe theoretical models, of clearly defined baryogenesis, and of a well defined mechanism of phase transitions.

Correspondence: A. W. Beckwith:     projectbeckwith2@yahoo.com

PACS numbers**:** 03.75.Lm, 11.27.+d, 98.65.Dx, 98.80.Cq, 98.80.-k

# I. INTRODUCTION

As of June 2005, an effort was made to combine reconstruction of data gathering techniques with the requirement of the JDEM dark matter-dark energy search for the origins of dark matter in the early universe[1]. This has, among other things, lead to methodologies being presented which could shed light as to the initial formation of scalar potentials which could contribute to CMB background radiation. In doing so, it was noted that initial dimensions, as postulated by Quin, Pen, and Silk[2] presented evidence as to how three extra dimensions play a role in explaining how at very short distances gravity would have a $r^{-5}$ spatial behavior dependence in force calculations. We do believe that in the initial stages of cosmic inflation, that space, indeed had additional dimensions and that the dimensions play a role as far as nucleation of a new universe.

We have, therefore, written up how to reconstruct potentials, using the methodology presented by Kadota et al[3], but we also think that it is important to pick out properties of the potential in question with respect to early universe models, since CMB data as presently configured is too imprecise to get anything other than the standard FRW flat space metric, 1000 or so years after the big bang. So being the case, we have constructed a list of properties of what an early universe potential system, composed of di quark constituents in order to help researchers investigate CMB data more accurately for very early universe configurations.

# II. ORGANIZATION OF THE PAPER

**Appendix I, parts A and B** highlights what can be said about typical data reconstruction for early universe potential systems. **Appendix II** initiates setting up three regimes for an evolving potential model leading to chaotic inflation, in line with Dr.

Guths quartic potential system. **Appendix III** presents what is done with instanton models of what is called in the literature, QCD balls[4], for initially stable di quark pairs which we believe are the building blocks for a false vacuum nucleation of initial baryonic states of matter. Afterwards, in the main text of this document, I examine the break down of what Bunyi and Hsu of the U. of Oregon call a semi classical approximation[5], but which I call a consistency condition for di quark pair contributions to Guth style chaotic inflation. In making this consistency evaluation, I then refer to in **Appendix IV, part A and B** how and why the initial wave functionals used in forming this semi classical evaluation are formed. This uses the results of two accepted world press scientific articles, one published in IJMPB [6], and also another accepted already for publication in Modern Physics Letters B [7], which describe necessary and sufficient conditions for a false vacuum based construction of Gaussian wave functionals, which then have, due to the very short distances involved, a discrete state presentation which is then used in an inner product evaluation of potential systems.

It is interesting to note that the semi classical (consistency) condition so outlined in the main text works best initially for a modified driven washboard potential system, which is integrated over six dimensions, in line with Silks presentation as to the importance of higher dimensions being very significant for extremely small spatial dimensions. This is given more structure in **Appendix V**. I do believe that this is no accident, and is congruent with the Calabi Yau conjecture in string theory with the curling up of higher spatial dimensions, which after a certain phase in inflation no longer contribute significantly to the chaotic inflation paradigm presented by Guth[8].

How does this bend in with more observational techniques as given by astrophysics researchers? Early universe nucleation is too small a region of space for typical action integral arguments to be effectual. So I have presented an alternative, as given in **Appendix VII**, which I do believe is able to give a new structure as to how to consider the flux of particles from a cosmic nucleation stand point[9]. In addition, the initial configuration of matter states would not be treatable by the Einstein cosmological constant. But that the evolution of di quark states can, after the onset of inflation, evolve into an Einstein cosmological constant dominated epoch, as given by an argument based upon a modified Scherrer k essence argument. This argument is important, and is in the main text of this document. All of this has been presented in PANIC 2005, and will be included in the AIP proceedings of that conference[10].

This last step depends upon a break down of a thin wall approximation. I do believe that this is consistent with the quantum fluctuations of momentum discussed in the paper written by R. Aloisio[11] et al, about deformed special relativity and its relations to a supposed quantum gravitational background.

Finally, I make direct connection with Venzianos[12] postulates as to the links between Planck scale length, a scalar field term, and a wavelength approximately in sync with the initial scale of a nucleating universe. I suggest here that the initial cosmic nucleation diameter was of the order of Planck Length, and subsequently radically expanded afterwards, in a result consistent with cosmic inflation.

The initial impetus for making this effort was due to the following conundrum. As is commonly known in cosmology circles, one would expect a flat Friedman – Walker universe after 60 e-foldings, but beforehand one could expect sharp deviations as to flat space geometry. The moment one would expect to have deviations from the flat space geometry would closely coincide with Rocky Kolb's model for when degrees of freedom would decrease from over 100 degrees of freedom to roughly ten or less during an abrupt QCD phase transition[13]. As was mentioned by Joe Lykken , the CMB model should yield a distinct 'signal' which is lending toward a non flat cosmological metric space potential which can be seen to be initiating a phase transition at about the end of the 60 e-folding regime of cosmological expansion[14]. My own model is useful for such QCD phase transitions; while Kenji Kadoka's potential reconstruction scheme is not specific as to a **<u>UNIQUE</u>** potential structure. It would be enough in itself to try to combine the two techniques as to go before the thousand year mark Kenji mentioned as to data sets permitting potential reconstruction, and to find evidence as to CMB background as to the initial phases of CMB generation leading to the datum Kolb mentioned as to the decrease in cosmic microwave radiation to its present value as a result of a QCD phase transition in the expansion of the early universe.

## III. BRIEF RE CAP OF QINS EXTRA DIMENSIONS FROM DARK MATTER ARTICLE, PLUS THE EVOLVING POTENTIAL SYSTEM ACCOMODATING DI QUARK SCALAR FIELDS

As mentioned, Quinn's article[2] gives a new force law, with respect to distances at or below 1$nm$ in length. As presented in the article [10], this appears to be a verification of the existence of small but non infinitesimal extra dimensions. The key assumption which was used in their paper was a force law of the general form for distances $r << R$ :

$$F = \alpha \cdot \frac{GMm}{r^{2+n}} \tag{3.1}$$

Here, $\alpha$ is a constant with dimensions $[length]^n$, G is the gravitational constant, and M and *m* are the masses of the two particles and. $\alpha \equiv R^n$ was set, while the value of *n* was, partly to fit with an argument given by Volt and Wannier[15] that the quantum mechanical cross section for collision is twice the corresponding classical value, if one assumes a central force field dependence of $r^{-5}$ This all together, if one assumes that initially *r* is of the order of magnitude of Planck's length $l_P$ would lead to extremely strong pressure values upon the domain walls of a nucleated scalar field initial states, which I claim would lead to a quite necessary collapse of the thin wall approximation. This collapse of the thin wall approximation set the stage for an Einstein constant dominated regime in inflation, if one adheres to a version of Scherrer's K essence theory[16] results for modeling the di quark pairs used as an initial starting point for soliton-anti soliton pairs(S-S') in the beginning of quantum nucleation of our universe.[10]

We should note that **Appendix I** as given gives a necessary and sufficient condition for constructing a potential system, initially in the false vacuum mode of potential, due to a pop up of a di quark state[10]. Here, for reasons of scale, we set $M_P$ as a Planck mass, and the 2nd mass, m, as considerably smaller. The scalar field term $\phi$ is constructed in terms of di quarks, in line with the soliton- anti soliton (S-S') used in the two accepted articles using similar constructions in IJMPB, etc [6,7], and $\phi^*$ is, here, picked in terms of the limits of quantum fluctuations of a scalar field, in line with Guths model of chaotic inflation[17]. Furthermore we write the potentials $V_1$, $V_2$, and $V_3$ in terms of S-S' di quark pairs nucleating and then contributing to a chaotic inflationary scalar potential system[10].

$$V_1(\phi) = \frac{M_P^2}{2} \cdot (1 - \cos(\phi)) + \frac{m^2}{2} \cdot (\phi - \phi^*)^2 \qquad (3.2a)$$

$$V_2(\phi) \approx \frac{(1/2) \cdot m^2 \phi^2}{(1 + A \cdot \phi^3)} \qquad (3.2b)$$

$$V_3(\phi) \approx (1/2) \cdot m^2 \phi^2 \qquad (3.2c)$$

The difference between these potentials becomes extraordinarily important in considering how the nucleating universe system evolves in time from the onset of the big bang itself. Furthermore, as a convenience, I have bench marked the $\phi^*$ term via the following procedure. We consider if and when we have classical and quantum fluctuations approximately giving the same value for a phase value of [17]

$$\phi^* \equiv \left(\frac{3}{16 \cdot \pi}\right)^{\frac{1}{4}} \cdot \frac{M_P^{3/2}}{m^{\frac{1}{2}}} \cdot M_P \to \left(\frac{3}{16 \cdot \pi}\right)^{\frac{1}{4}} \cdot \frac{M_P^{3/2}}{m^{\frac{1}{2}}} \qquad (3.3)$$

where we have set $M_P$ as the typical Planck mass which we normalized to being unity in this paper for the hybrid false vacuum – inflaton field cosmology example, as well as having set the general evolution of our scalar field as having the form of [17]

$$\phi \equiv \tilde{\phi}_0 - \frac{m}{\sqrt{12 \cdot \pi \cdot G}} \cdot t \qquad (3.4)$$

This then permits us to look at how consistent having a di quark model, with a thin wall approximation is, with the evolving potential system, given above. It is important, since we are finding that having the additional dimensions specified in the beginning permits us to have a more physically consistent picture of how the phase transition to an Einstein constant dominated cosmology occurs in the first place. This is especially relevant from going from the 1st to the 3rd potential given above. **Appendix III** presents what is done with instanton models of what is called in the literature, QCD balls[4], for initially stable di quark pairs, which is what we are assuming with this construction of $\phi$, and we can use to obtain S-S' type pairs which are then used to construct wave functional representation of early universe states. This is in part based upon **Appendix IV, part A and B** on how and why the initial wave functional used in forming this semi classical evaluation are formed. This uses the results of two accepted world press scientific articles, one published in IJMPB [6], and also another accepted already for publication in Modern Physics Letters B [7]. The important thing to consider here, though is that we are looking at understanding the existence of the phase transformation from the first to the third potential occurs, and what it says about the formation of conditions relevant toward an Einstein constant dominated cosmology

## IV: CRITERIA USED BY BUNYI AND HSU, WHICH WE CALL A CONSISTENCY CONDITION REFLECTING THE OCCURANCE OF A PHASE TRANSITION.

Let us first consider an elementary definition of what constitutes a semi classical state. As visualized by Buniy and Hsu,[5] it is of the form $|a\rangle$ which has the following properties:

i) Assume $\langle a|1|a\rangle = 1$

(Where 1 is an assumed identity operator, such that $1|a\rangle = |a\rangle$)

ii) We assume that $|a\rangle$ is a state whose probability distribution is peaked about a central value, in a particular basis, defined by an operator $Z$

a) Our assumption above will naturally lead, for some *n* values

$$\langle a|Z^n|a\rangle \equiv (\langle a|Z|a\rangle)^n \tag{4.1}$$

Furthermore, this will lead to, if an operator $Z$ obeys Eq. (4.1) that if there exists another operator, call it $Y$ which does not obey Eq. (4.1), that usually we have non commutativity

$$[Y,Z] \neq 0 \tag{4.2}$$

Buniy and Hsu[5] speculate that we can, in certain cases, approximate a semi classical evolution equation of state for physical evolution of cosmological states with respect to classical physics operators. This well may be possible for post inflationary cosmology; however, in the initial phases of quantum nucleation of a universe, it does not apply. We do this with a potential system, with S-S' di quark constituents we model via using[10]

$$\phi \equiv \pi \cdot [\tanh b(x - x_a) + \tanh b(x_b - x)] \tag{4.3}$$

We can, in this give an approximate wave function as given by a discretized version of the wave functional given for the first potential system as in Appendix IV, B:

$$\psi \cong c_1 \cdot \exp(-\tilde{\alpha} \cdot \phi(x)) \tag{4.4}$$

Then we can look to see if we have[5,10]

$$\left( \int_{x_a}^{x_b} \psi \cdot V_i \cdot \psi \cdot 4\pi \cdot x^2 \cdot dx \right)^N \equiv \int_{x_a}^{x_b} \psi \cdot [V_i]^N \cdot \psi \cdot 4 \cdot \pi \cdot x^2 \cdot dx \bigg|_{i=1,2,3} \tag{4.5}$$

This was later generalized, in the initial phases of nucleation for the 1$^{st}$ potential system as being, in six initial dimensions

$$\left( \int_{x_a}^{x_b} \psi \cdot V_i \cdot \psi \cdot 4\pi \cdot x^5 \cdot dx \right)^N \equiv \int_{x_a}^{x_b} \psi \cdot [V_i]^N \cdot \psi \cdot 4 \cdot \pi \cdot x^5 \cdot dx \bigg|_{i=1} \tag{4.6}$$

In addition, the analysis of how to work with a ratio of the values of the left and right hand sides of eq (4.5 and (4.6)as a way of looking at the consistency of what has been called the semi classical approximation would lead to analyzing

$$\Phi_{i,n,N}(ratio) \equiv \left. \frac{\left( \int_{x_a}^{x_b} \psi \cdot V_i \cdot \psi \cdot x^{2+n} \cdot dx \right)^N}{\int_{x_a}^{x_{bf}} \psi \cdot [V_i]^N \psi \cdot x^{2+n} \cdot dx} \right|_{i=1,2,3} \qquad (4.7)$$

The first coefficient, i, denotes which potential system is picked, and ranges in value from 1 to 3. The second coefficient, $n$, is either 0 or 3, depending upon what dimensionality is assumed for this problem. The third coefficient, N, is freely ranging in values from 1 up to 100. I as a convenience often worked with N= 9. This eventually led to the calculations of **Appendix V**, which highlight the importance of higher dimensionality in the initial stages of nucleation, for the first potential system.

Assuming that this is a valid initial dimensional approximation, we did the following for the three potentials.

a. Assumed that the scalar wave functional term was decreasing in '*height*' and increasing in '*width*' as we moved from the first to the third potentials. $\phi$ also had a definite evolution of the domain wall from a '*near perfect*' thin wall approximation to one which had a considerable slope existing with respect to the wall.

b. We also observed that in doing this sort of model that there was a diminishing of magnitude from unity for Eq. (4.7) for large *N* values, regardless if or not the thin wall approximation was weakened as we went from the first to the third potential system. In doing to, we also noted that even in Eq. (4.7) for the first potential, Eq. (4.7) had diminishing applicability as a result for decreasing $b$ values in Eq. (4.3), which corresponded to when the thin wall approximation was least adhered to.

We also observed that for the third potential, that there was never an overlap in value between the left and right hand sides of Eq. (4.5) and Eq. (4.6), regardless of whether the thin wall approximation was adhered to. In other words, the third potential was least linkable to a semi classical approximation of physical behavior linkable to a physical system, while Eq. (4.5) and Eq. (4.6) worked best for a thin domain wall approximation to Eq. (4.4) in the driven sine Gordon approximation of a potential system. In all this, we assumed that the small perturbing term added to the $(1-\cos(\phi))$ part of Eq. (3.2a) was a physical driving term to a very classical potential system $(1-\cos(\phi))$ which had a quantum origin consistent with the interpretation of a false vacuum nucleation of the sort initially formulated by Sidney Coleman.[18] Furthermore, as we observed an expanding 'width' in Eq. (4.3), the alpha term in Eq (4.4) shrank in its value, corresponding to a change in the position of constituent S-S' components in the scalar field given in this model. The S-S' terms roughly corresponded to di quark pairs.

  c. Chaotic inflation in cosmology is, in the sense a quartic potential portrayed by Guth,[17] a general term for models of the very early Universe which involve a short period of extremely rapid (exponential) expansion; blowing the size of what is now the observable Universe up from a region far smaller than a proton to about the size of a grapefruit (or even bigger) in a small fraction of a second. This process smoothes out space-time to make the Universe flat, but is not in the model presented linkable in the chaotic inflationary region given by the third potential to any semi classical arguments. The relative good fit of Eq. (4.7) for the first potential is in itself an argument that the thin wall approximation breaks down past the point of baryogenesis after the chaotic inflationary regime is initiated by the third potential as modeled by Guth.[17]

Since we have established this, we should then attempt to consider if the higher dimensional physical state relevant to the 1st potential system are countable. Yes they are, but not by ordinary least action principal arguments[19]. I give a variant of what could be analyzed in **Appendix VII**, after stating that the earlier least action counting algorithms referenced in **Appendix VI** (summary of Garrigas work)[20] is not germane to such a small scale physical system. This then leads to to consider what the evolving state of di quark pairs says about , from a Scherrer k essence stand point of how the evolution to Guth chaotic inflation, as given by the third potential corresponds to the rise of an Einstein constant dominated inflationary cosmology[10].

## V. HOW DARK MATTER TIES IN, USING PURE KINETIC *K* ESSENCE AS DARK MATTER TEMPLATE FOR A NEAR THIN WALL APPROXIMATION OF THE DOMAIN WALL FOR $\phi$

We define k essence as any scalar field with non-canonical kinetic terms. Following Scherrer,[10,21,22] we introduce a momentum expression via

$$p = V(\phi) \cdot F(X) \tag{5.1}$$

where we define the potential in the manner we have stated for our simulation as well as set[10,21,22]

$$X = \frac{1}{2} \cdot \nabla_\mu \phi \; \nabla^\mu \phi \tag{5.2}$$

and use a way to present F expanded about its minimum and maximum[10,21,22]

$$F = F_0 + F_2 \cdot (X - X_0)^2 \tag{5.3}$$

where we define $X_0$ via $F_X|_{X=X_0} = \dfrac{dF}{dX}\bigg|_{X=X_0} = 0$, as well as use a density function[10,21,22]

$$\rho \equiv V(\phi) \cdot [2 \cdot X \cdot F_X - F] \tag{5.4}$$

where we find that the potential neatly cancels out of the given equation of state so[10,21,22]

$$w \equiv \frac{p}{\rho} \equiv \frac{F}{2 \cdot X \cdot F_X - F} \tag{5.5}$$

as well as a growth of density perturbations terms factor Garriga and Mukhanov[20] wrote as

$$C_x^2 = \frac{(\partial p / \partial X)}{(\partial \rho / \partial X)} \equiv \frac{F_X}{F_X + 2 \cdot X \cdot F_{XX}} \tag{5.6}$$

where $F_{XX} \equiv d^2 F / dX^2$, and since we are fairly close to an equilibrium value, we pick a value of X close to an extremal value of $X_0$.[10,21,22]

$$X = X_0 + \tilde{\varepsilon}_0 \tag{5.7}$$

where, when we make an averaging approximation of the value of the potential as very approximately a constant, we may write the equation for the k essence field as taking the form (where we assume $V_\phi \equiv dV(\phi)/d\phi$)

$$(F_X + 2 \cdot X \cdot F_{XX}) \cdot \ddot{\phi} + 3 \cdot H \cdot F_X \cdot \dot{\phi} + (2 \cdot X \cdot F_X - F) \cdot \frac{V_\phi}{V} \equiv 0 \tag{5.8}$$

as approximately

$$(F_X + 2 \cdot X \cdot F_{XX}) \cdot \ddot{\phi} + 3 \cdot H \cdot F_X \cdot \dot{\phi} \cong 0 \tag{5.9}$$

which may be re written as[10,21,22]

$$(F_X + 2 \cdot X \cdot F_{XX}) \cdot \ddot{X} + 3 \cdot H \cdot F_X \cdot \dot{X} \cong 0 \tag{5.10}$$

In this situation, this means that we have a very small value for the growth of density pertubations[10,21,22]

$$C_S^2 \cong \frac{1}{1 + 2 \cdot (X_0 + \tilde{\varepsilon}_0) \cdot (1/\tilde{\varepsilon}_0)} \equiv \frac{1}{1 + 2 \cdot \left(1 + \frac{X_0}{\tilde{\varepsilon}_0}\right)} \tag{5.11}$$

when we can approximate the *kinetic energy* from

$$(\partial_\mu \phi) \cdot (\partial^\mu \phi) \equiv \left(\frac{1}{c} \cdot \frac{\partial \phi}{\partial \cdot t}\right)^2 - (\nabla \phi)^2 \cong -(\nabla \phi)^2 \rightarrow -\left(\frac{d}{dx}\phi\right)^2 \tag{5.11a}$$

and, if we assume that we are working with a comparatively small contribution w.r.t. time variation but a very large, in many cases, contribution w.r.t. spatial variation of phase

$$|X_0| \approx \frac{1}{2} \cdot \left(\frac{\partial \phi}{\partial x}\right)^2 \gg \tilde{\varepsilon}_0 \tag{5.11b}$$

$$0 \leq C_S^2 \approx \varepsilon^+ \ll 1 \tag{5.12}$$

And [10,20]

$$w \equiv \frac{p}{\rho} \cong \frac{-1}{1 - 4 \cdot (X_0 + \tilde{\varepsilon}_0) \cdot \left( \frac{F_2}{F_0 + F_2 \cdot (\tilde{\varepsilon}_0)^2} \cdot \tilde{\varepsilon}_0 \right)} \approx 0 \tag{5.13}$$

We get these values for the phase $\phi$ being nearly a box, i.e. the thin wall approximation for $b$ being very large in Eq. (4.3); this is consistent with respect to Eq. (5.13) main result, with $w \equiv \frac{p}{\rho} \cong 0 \Rightarrow$ treating the potential system given by the first potential (modified sine Gordon with small quantum mechanical driving term added) as a semi classical system leading to Eq. (4.7) nearly being unity. This also applies to the formation of S-S' pair formation due to the di quarks as alluded to in Zhitinisky's[16] formulation of QCD balls with an axion wall squeezer having a 'thin wall' character.

When we observed

$$|X_0| \approx \frac{1}{2} \cdot \left( \frac{\partial \phi}{\partial x} \right)^2 \cong \frac{1}{2} \left[ \delta_n^2(x + L/2) + \delta_n^2(x - L/2) \right] \tag{5.14}$$

with

$$\delta_n(x \pm L/2) \xrightarrow[n \to \infty]{} \delta(x \pm L/2) \tag{5.15}$$

as the slope of the S-S' pair approaches a box wall approximation in line with thin wall nucleation of S-S' pairs being in tandem with $b \to$ *larger*. Specifically, in our simulation, we had $b \to 10$ above, rather than go to a pure box style representation of S-S' pairs; this could lead to an unphysical situation with respect to delta functions giving infinite values of infinity, which would force both $C_s^2$ and $w \equiv \frac{p}{\rho}$ to be zero for

$$|X \approx X_0| \cong \frac{1}{2} \cdot \left(\frac{\partial \phi}{\partial x}\right)^2 \to \infty$$ if the ensemble of **S-S'** pairs were represented by a pure thin wall approximation,[20] i.e., a box. If we adhere to a finite but steep slope convention to modeling both $C_s^2$ and $w \equiv \frac{p}{\rho}$, we get the following: When $b \geq 10$ we obtain the conventional results of

$$w \cong \frac{-1}{1 - 4 \cdot \frac{X_0 \cdot \tilde{\varepsilon}_0}{F_2}} \to -1 \tag{5.16}$$

and recover Scherrer's solution for the speed of sound [10,21,22]

$$C_S^2 \approx \frac{1}{1 + 4 \cdot X_0 \left(1 + \frac{X_0}{2 \cdot \tilde{\varepsilon}_0}\right)} \to 0 \tag{5.17}$$

(If an example $F_2 \to 10^3, \tilde{\varepsilon}_0 \to 10^{-2}, X_0 \to 10^3$). Similarly, we would have if $b \to 3$ in Eq. (4.5)

$$w \cong \frac{-1}{1 - 4 \cdot \frac{X_0 \cdot \tilde{\varepsilon}_0}{F_2}} \to -1 \tag{5.18}$$

and

$$C_S^2 \approx \frac{1}{1 + 4 \cdot X_0 \left(1 + \frac{X_0}{2 \cdot \tilde{\varepsilon}_0}\right)} \to 1 \tag{5.19}$$

if $F_2 \to 10^3$, $\tilde{\varepsilon}_0 \to 10^{-2}$. Furthermore $|X_0| \to$ *a small value*, which for $b \to 3$ in Eq. (5) would lead to $C_s^2 \approx 1$, i.e., when the wall boundary of a S-S' pair is no longer approximated by the thin wall approximation. This eliminates having to represent the initial state as behaving like pure radiation state (as Cardone[23] postulated), i.e., we then recover the cosmological constant. When $|X_0| \approx \frac{1}{2} \cdot \left(\frac{\partial \phi}{\partial x}\right)^2 \gg \tilde{\varepsilon}_0$ no longer holds, we can have a hierarchy of evolution of the universe as being first radiation dominated, then dark matter, and finally dark energy.

If $|X \approx X_0| \cong \frac{1}{2} \cdot \left(\frac{\partial \phi}{\partial x}\right)^2 \to \infty$, neither limit leads to a physical simulation that makes sense; so, in this problem, we then refer to the contributing slope as always being large but not infinite. We furthermore have, even with $w = -1$

$$C_s^2 \equiv 1 \xrightarrow[b1 \to 3]{} 1 \qquad (5.20)$$

indicating that the evolution of the magnitude of the phase $\phi \to \varepsilon^+$ corresponds with a reduction of our cosmology from a dark energy dark matter mix to the more standard cosmological constant models used in astrophysics. This coincidently is when the semi classical evaluation involving S-S' di quark pairs breaks down, as given by Eq. (4.7) being much smaller than unity and corresponds to the b of Eq. (4.3) for $\phi \to \varepsilon^+$ being quite small. It also denotes a region where there is a dramatic reduction of the degrees of freedom of the FRW space time metric, as Kolb postulated[24,25] so that we can then visualize cosmological dynamics being governed by the Einstein constant at the conclusion of the cosmological inflationary period

## VI: CONCLUSION

Veneziano's model [12] gives us a neat prescription of the existence of a Planck's length dimensionality for the initial starting point for the universe via:

$$l_P^2/\lambda_S^2 \approx \alpha_{GAUGE} \approx e^\phi \tag{6.1}$$

where the weak coupling region would correspond to where $\phi \ll -1$ and $\lambda_S$ is a so called quanta of length, and $l_P \equiv c \cdot t_P \sim 10^{-33} cm$. As Veneziano implies by his 2nd figure [6], a so called scalar dilaton field with these constraints would have behavior seen by the right hand side of his figure one, with the $V(\phi) \to \varepsilon^+$ but would have no guaranteed false minimum $\phi \to \phi_F < \phi_T$ and no $V(\phi_T) < V(\phi_F)$. The typical string models assume that we have a present equilibrium position in line with strong coupling corresponding to $V(\phi) \to V(\phi_T) \approx \varepsilon^+$ but no model corresponding to potential barrier penetration from a false vacuum state to a true vacuum in line with Coleman's presentation.[5,20] However, FRW cosmology[26] will in the end imply

$$t_P \sim 10^{-42} \sec onds \Rightarrow size\ of\ universe \approx 10^{-2} cm \tag{6.2}$$

which is still huge for an initial starting point, whereas we manage to in our S-S' 'distance model' to imply a far smaller but still non zero radii for the initial 'universe' in our model.

We find that the above formulation in Eq. (6.1) is most easily accompanied by the given S-S' di quark pair basis for the scalar field used in this paper, and that it also is consistent with the initial scalar cosmological state evolving toward the dynamics of the cosmological constant via the *k* essence argument built up near the end of this document.

Furthermore, we also argue that the semi classical analysis of the initial potential system as given by Eq. (4.7) and its subsequent collapse is de facto evidence for a phase transition to conditions allowing for CMB to be created at the beginning of inflationary cosmology.

We are fortunate as shown in **Appendix V** that for determining the relative good fit of Eq. (4.7) that the relative domain walls slope of the initial phase given by Eq. (4.5) was not terribly significant, for the first potential system, which dove tails with Eq. (4.1) merely pushing out the domain walls, as a primary effect, for a driven sine Gordon type modeling of false vacuum nucleation. As ,mentioned earlier, this was actually heightened by the extra dimensionality as alluded to by the power law relationship in Eq. (4.1) making an almost perfect equality between the left and right hand sides of Eq. (4.7). That the ratio Eq. (4.7) in **Appendix V** had varying values, showing different degrees of break down of this relationship for the $2^{nd}$ transitional potential, due to differences in dimensionality and slope of the scalar field as given by Eq. (4.3) is probably due to this representing the abrupt loss of numbers of degrees of freedom Rocky Kolb has mentioned as part of a phase transition. Needless to say though, as we evolve toward the Einstein cosmological constant era and chaotic inflation, as given by the $3^{rd}$ potential, we should keep in mind very real limits as to the comparative sharpness of the slope of the scalar field as given by Eq. (4.3)

K essence analysis argues against making $b$ in Eq. (4.3) too large, i.e., if we have a 'perfect' thin wall approximation to our S-S' di quark pairs, we will have the unphysical speed of sound results plus other consequences detailed in the k essence section of the document which we do not want. On the other hand, the semi classical analysis brought up in the section starting with Eq. (4.5), Eq. (4.6) and summarized by Eq. (4.7) shows us that a close to the thin wall approximation for S-S' di quark pairs gives an optimal fit for consistency in the potential with the wave functions exhibiting a thin wall approximation 'character'. It is useful to note that our kinetic model can be compared with the very interesting Chimentos [27] purely kinetic k –essence model, with density fluctuation behavior at the initial start of a nucleation process. The model indicate our density function reach $\rho$ = constant after passing through the tunneling barrier as mentioned in our nucleation of a S-S' pair ensemble. This is when the Einstein constant becomes dominant and that the semi classical approximation in Eq. (4.7) for a domain wall at the time the comparative thin wall approximation S-S' pair ceases to be relevant.

Our initial attempt here very likely should be re visited, especially if the sort of brane world objects referred to by Trodden et a[28] are used in a future calculation for initial nucleation states. However, this should all be done to re calibrate how to fill in the CMB contribution toward reconstruction of a suitable class of potentials which could shed light not only on the origins of baryogenesis, in early universe models, but also in determining how dark matter-dark energy could contribute to the formation of initial inflationary cosmology parameters. The hope is that if suitable data reconstruction methodology is obtained and refined, that one could as an example determine how the initial physical fundamental constants could be set as they are, as well understand how dark matter-dark energy contribute to the initial origins of CMB itself This also would allow us to improve upon the particle flux from nucleation argument we used, using Gongs [29] approximate construction in order to get around limitations in least action principles due to tiny spatial dimensions.

**[Insert figures 1a, 1b, and then figures 2a, 2b *with captions* here]**

Furthermore, we should note that these nucleation configurations fit in well with the following model of false vacuum nucleation.

**[Insert figure 3 *with caption* here]**

This is in line with the first specified potential as given in Eq. (3.2a) which we claim eventually becomes in sync with Eq. (3.2c). Further progress in investigating this phenomenology should take into account the datum so mentioned in the text, about the original multiple dimensions in the initial phases of a nucleating universe, which subsequently are reduced as the scalar potential evolves toward the chaotic potential given in Eq. (3.2c). This should permit us to be able to reconstruct potentials far closer to the big bang itself, than the 1000 or so year limit alluded to by Dr. Kadota in his May 2005 Pheno talk given in Madison, Wisconsin[30]. This in its own way will entail considerable additional analytical work, along the lines first specified by *Edmund J. Copelan et al.* in their ground breaking tome on potential reconstruction techniques applied to cosmology[31] .It is worthwhile to note that the orientation of my white paper was in unifying certain techniques, and methodology of what is known in the literature as QCD balls in an instanton configuration to use data reconstruction in order to obtain information on dark matter physics. In doing so, I wound up using a lot of ideas, as was done by other physicists considering early universe nucleation models, from condensed matter physics. The emphasis though of the presented concept was in setting up a template as to examine what actually constitutes dark matter. This should be what future inquiry should be directed toward.

# Figure captions

**Fig 1a,b:** Evolution of the phase from a thin wall approximation to a more nuanced thicker wall approximation with increasing L between S-S' instanton components. The 'height' drops and the 'width' L increases correspond to a de evolution of the thin wall approximation. This is in tandem with a collapse of an initial nucleating 'potential' system to the standard chaotic scalar $\phi^2$ potential system of Guth. As the 'hill' flattens, and the thin wall approximation dissipates, the physical system approaches standard cosmological constant behavior.

**Fig 2a,b:** As the walls of the S-S' pair approach the thin wall approximation, a normalized distance, $L=9 \rightarrow L=6 \rightarrow L=3$, approaches delta function behavior at the boundaries of the new nucleating phase. As $L$ increases, the delta function behavior subsides dramatically. Here, the $L=9 \Leftrightarrow$ conditions approaching a cosmological constant. $L=6 \Leftrightarrow$ conditions reflecting Scherrer's dark energy-dark matter mix. $L=3 \Leftrightarrow$ approaching unphysical delta function contributions due to a pure thin wall model.

**Fig 3:** Initial configuration of the domain wall nucleation potential as given by Eq. (4.4a) which we claim eventually becomes in sync with Eq. (4.4c) due to the phase transition alluded to by Dr. Edward Kolbs model of how the initial degrees of freedom declined from over 100 to something approaching what we see today in flat Euclidian space models of space time (i.e. the FRW metric used in standard cosmology)

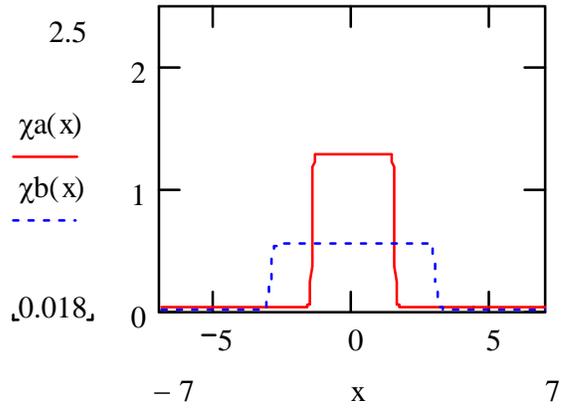

**Figure 1a, 1b**

**Beckwith**

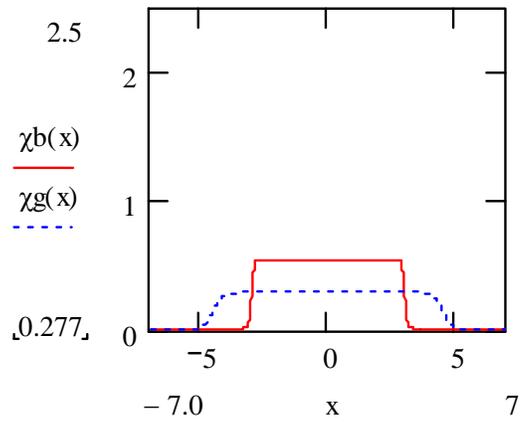

**Figure 2a, 2b**

**Beckwith**

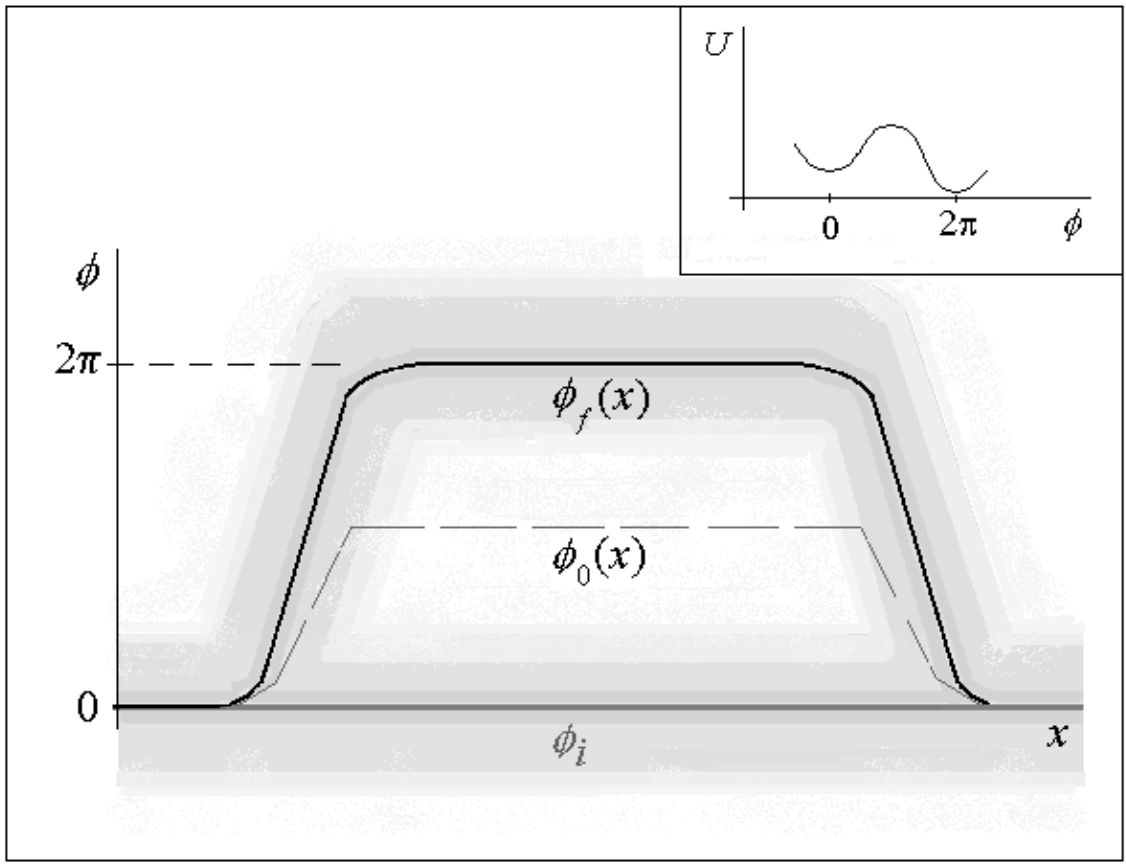

**Figure 3**

**Beckwith**

# APPENDIX IA: INITIAL STATEMENT OF KADOTAS POTENTIAL RECONSTRUCTION METHODOLOGY

Kenji Kadot*a* of FNAL in Pheno 2005 [30] and also in arXIV [3] talked of comparing two graphs, one with a combination of scalar potential terms $\left[ 3 \cdot \left( \frac{V'}{V} \right)^2 - 2 \cdot \frac{V''}{V} \right]$ against $[(\xi)]$ (Mpc) with a graph of

$m(j)$ against **_mode numbers_**. Here, in this situation,

$$m(j) = \text{linear combination of } \{P(j)\} \tag{1a}$$

And when we set $t_{END}$ = the demarcation of the end of time for the inflation, for a scale factor $a$ leads to

$$\xi \equiv -\int_{t}^{t_{END}} \frac{dt'}{a(t')} \tag{1b}$$

In this situation, the $\{P(j)\}$ refer to pixel data slices which show up in

$$\left[ 3 \cdot \left( \frac{V'}{V} \right)^2 - 2 \cdot \frac{V''}{V} \right] \equiv \sum_i p_i \cdot B_i (\ln \xi) \tag{2}$$

We should identify the left hand side of equation 2 with the derivative of a function $G(\xi)$, i.e.

$$\frac{dG(\xi)}{d\xi} \equiv \sum_i p_i \cdot B_i (\ln \xi) \tag{2a}$$

This is when Kadota et al defined

$$B_i (\ln \xi) = \begin{cases} 1 \\ 0 \end{cases} \text{ with a value of 1 iff } \ln \xi_i < \ln \xi < \ln \xi_{i+1} \tag{3}$$

In the most recent arXIV article, Kadota defined a procedure as to how to identify useful entries as to acceptable $\{P(j)\}$ values as to a simplified scalar potential structure which is

$$V(\phi) \equiv \left(V_0 \cdot e^{\lambda \cdot (\phi - \phi_0)}\right) \cdot \left[1 + c \cdot e^{-\nu \cdot (\phi - \phi_0)^2}\right]$$ for a perturbation centered at $\phi \equiv \phi_0$ where this has $|\lambda|, c \ll 1, \ |\nu| \gg 1$ , so then after Kadota defined

$$\phi \cong \lambda \cdot \ln \xi \tag{4}$$

so one could write

$$\phi_0 \equiv \lambda \cdot \ln \xi_0 \tag{5}$$

He, Kadota, obtained graphical behavior as seen in his fig 8 and fig 9 of his arXIV article[3]. An even simpler situation graphically emerged when Kadota set the left hand side of Eq. (1a) equal to a constant which permitted him, using Eq. ( 2 ) and Eq. (3) above to give constant values to the $p_i$ pixels, which was equivalent to his figure 7 which was for a potential system leading to a constant spectral index value, n when he defined via linking n – 1 to the derivative with respect to k of an expression of the primordial power spectrum $\mathrm{P}(k)$ via

$$n - 1 = \frac{d\mathrm{P}(k)}{dk} \tag{6}$$

Here, in this situation we have that if we interpret $\vartheta(1)$ as an order of magnitude constant of about $1 < \vartheta(1) < 10$. We should also note that often $\vartheta(1)$ is often set very close to 1 itself.

$$k = \vartheta(1) \cdot a \cdot H \equiv a \cdot H \tag{7}$$

The exact particulars of the power spectra $\mathrm{P}(k)$ are in Kadoka's well written arXIV paper, but it suffices to say that the natural logarithm of the power spectra $\mathrm{P}(k)$ is equal to an integral over $\xi$ values from zero to infinity, with part of the integrand involving a so called 'window function' times the power spectra $P(k)$, for $G(k)$ of equation 2.2a above. I do believe one can say the following:

Kenji Kadoka's methodology permits the general reconstruction of potentials as up to about 1000 years after the big bang. The issue at stake though is if or not re constructive methodology using some of these same methods could be countenanced going up to the end of the 60 e-folding period commonly viewed as the demarcation between flat and curved space, with a curved space milieu being the regime of active nucleation of our universe. This would entail, among other things, finding traces in CMB data of the initial signature of the big bang itself and tying it into a QCD style phase transition.

## APPENDIX I B: USING JDEM ANALYSIS OF DATA WITH THIS BUILT UP POTENTIAL SYSTEM, WITH KADOTAS POTENTIAL RECONSTRUCTION PROCEDURES

The first step would be to refine the analytical algorithms to, give reliable data inputs into the right hand side [3,30] of $\frac{dG(\xi)}{d\xi} \equiv \sum_i p_i \cdot B_i(\ln \xi)$, where the left hand side of this equation actually could use, in a modified format the procedure given in Eq (3.2a) to Eq (3.2c) of the main text, and this done to obtain a match up of the acceptable $p_i$ entries with CMB data.

This would entail use of Monte Carlo simulations as well as far more developed analysis of how to obtain acceptable $p_i$ entries in a more realistic manner than the toy problem analyzed by *Kadoka's* toy problem[3,30] example which he presented in fig 7 of his arXIV article[3].

Afterwards, once acceptable procedures are outlined as to finding acceptable $p_i$ entries for potentials other than the potential given by Kadoka's test scalar potential given as

$$V(\phi) \equiv \left(V_0 \cdot e^{\lambda \cdot (\phi - \phi_0)}\right) \cdot \left[1 + c \cdot e^{-\nu \cdot (\phi - \phi_0)^2}\right] \tag{1}$$

The potential reconstruction I believe could be greatly aided by some of the initial effective contributions of extra dimensionality and of side effects of the baryogenesis mentioned in the formation of our early universe potential nucleation model The idea would be to find ways to obtain data sets via techniques most congruent to reliable potential reconstruction of the early inflationary cosmos. Before the 1000 or so year limit specified by Kenji Kadota in discussions I had with him at Pheno 2005 [30].

If finding acceptable match up of data sets with how to reconstruct a complicated potential beyond the one given by Eq (1) above was completed in general. Then one would face a discussion with manufacturers of the satellite used for dark matter searching as to tailor made electronics which would be acceptable for obtaining sufficient data sets. I am assuming that this investigation would be one out of many being used in the upcoming satellite mission.

# APPENDIX II: LINKS TO THE POTENTIAL SYSTEM USED FOR COSMOLOGICAL NUCLEATION

$$\begin{aligned} V_1 &\to V_2 &\to V_3 \\ \phi(increase) \leq 2\cdot\pi &\to \phi(decrease) \leq 2\cdot\pi \to \phi \approx \varepsilon^+ \\ t \leq t_P &\to t \geq t_P + \delta\cdot t \to t \gg t_P \end{aligned} \qquad (1)$$

We described the potentials $V_1$, $V_2$, and $V_3$ in terms of S-S' di quark pairs nucleating and then contributing to a chaotic inflationary scalar potential system.

$$V_1(\phi) = \frac{M_P^2}{2}\cdot(1-\cos(\phi)) + \frac{m^2}{2}\cdot(\phi-\phi^*)^2 \qquad (2a)$$

$$V_2(\phi) \approx \frac{(1/2)\cdot m^2 \phi^2}{(1+A\cdot\phi^3)} \qquad (2b)$$

$$V_3(\phi) \approx (1/2)\cdot m^2 \phi^2 \qquad (2c)$$

# APPENDIX III: INCLUDING IN NECESSARY AND SUFFICIENT CONDITIONS FOR FORMING A CONDENSATE STATE AT OR BEFORE PLANCK TIME $t_P$

For a template for the initial expansion of a scalar field leading to false vacuum inflationary dynamics in the expansion of the universe, Zhitnitsky's[4] formulation for how to form a condensate of a stable soliton style configuration of cold dark matter is a useful starting point for how an axion field can initiate forming a so called QCD ball. Zhitnitsky[4] uses quarks in a non-hadronic state of matter that, in the beginning, can be in di quark pairs. A di quark pair would permit making equivalence arguments to what is done with cooper pairs and a probabilistic representation as to find the relative 'size' of the cooper pair. We assume an analogous operation can be done with respect to di quark pairs. In doing so, calculations[4] for quarks being are squeezed by a so called QCD phase transition due to the violent collapse of an axion domain wall. The axion domain wall would be the squeezer to obtain a so called S-S' configuration. This presupposes a formation of a highly stable soliton type configuration in the onset due to the growth in baryon mass

$$M_B \approx B^{8/9} \tag{1}$$

This is due to a large baryon (quark) charge $B$ which Zhitnitsky[4] finds is smaller than an equivalent mass of a collection of free separated nucleons with the same charge. This provides criteria for absolute stability by writing a region of stability for the QCD balls dependent upon the inequality occurring for $B. > B_C$ (a critical charge value)

$$m_N > \frac{\partial M_B}{\partial B} \tag{2}$$

He[19] furthermore states that stability, albeit not absolute stability is still guaranteed for the formation of meta stable states occurring with

$$1 << B < B_C \tag{3}$$

If we make the assumptions that there is a balance between Fermi pressure $P_f$ and a pressure due to surface tension, with $\sigma$ being an axion wall tension value[4] so that

$$\left(P_\sigma \cong \frac{2\sigma}{R}\right) \equiv \left(P_f \cong -\frac{\Omega}{V}\right) \tag{4}$$

This pre supposes that $\Omega$ is some sort of thermodynamic potential of a non interacting Fermi gas, so that one can then get a mean radius for a QCD ball at the moment of formation of the value, when assuming $\tilde{c} \approx .7$, and also setting $B \approx B_C \propto 10^{+33}$ so that

$$R \equiv R_0 \cong \left( \frac{\tilde{c} \cdot B^{4/3}}{8 \cdot \pi \cdot \sigma} \right)^{1/3} \tag{5}$$

If we wish to have this of the order of magnitude of a Planck length $l_P$, then the axion domain wall tension must be huge, which is not unexpected. Still though, this pre supposes a minimum value of $B$ which Zhitnitsky[4] set as

$$B_C^{\exp} \sim 10^{20} \tag{6}$$

We need to keep in mind that Zhitnitsky[4] set this parameterization up to account for a dark matter candidate. I am arguing that much of this same concept is useful for setting up an initial condensate of di quark pairs as, separately S-S' in the initial phases of nucleation, with the further assumption that there is an analogy with the so called color super conducting phase (CS) which would permit di quark channels. The problem we are analyzing not only is equivalent to BCS theory electron pairs but can be linked to creating a region of nucleated space in the onset of inflation which has S-S' pairs. The S-S' pairs would have a distance between them proportional to distance mentioned earlier, $R_0$, which would be greater than or equal to the minimum Planck's distance value of $l_P$. The moment one would expect to have deviations from the flat space geometry would closely coincide with Rocky Kolb's model for when degrees of freedom would decrease from over 100 degrees of freedom to roughly ten or less during an abrupt QCD phase transition[4]. The QCD phase transition would be about the time one went from the first to the second potential systems mentioned above.

# APPENDIX IV A: WAVE FUNCTIONALS USED IN THIS MODEL AND THEIR ANALOGIES TO BLACK HOLE NUCLEATION

This idea of pair creation arose once again in a later context in an article by Dias and Lemos called 'Pair creation of black holes on a cosmic string background' where the so called 'amplitude' for the propagation from 'nothing' to a three dimensional surface boundary $\Sigma$ was given by the wave function [32] ( wave functional ):

$$\psi(h_{ij}, A_i) = \int d[g_{uv}] \cdot d[A_u] \cdot \exp(-I(g_{uv}, A_u)) \tag{1}$$

where $h_{ij}$ and $A_i$ are the induced metric and electromagnetic potential on the boundary $\Sigma = \partial M$ of a compact manifold $M$, and $I(g_{uv}, A_u)$ is the Euclidian action, with $d[g_{uv}]$ and $d[A_u]$ measures of the metric $g_{uv}$ and the Maxwell field $A_u$. Diaz and Lemos further state that a semi classical instanton approximation allows us to state that dominant contributions to the path integral come from metrics and Maxwell fields where are near the solutions which extremalize the Euclidian action and satisfy boundary conditions. So if we have this process, we may construct a wave function that that denotes the creation of a black hole via

$$\psi_{inst} \equiv B \cdot \exp(-I_{inst}) \tag{2}$$

where $B$ is a one loop contribution from quadratic fluctuations in the fields, $\delta^2 I$, and $I_{inst}$ is the classical action of the gravitational instanton that mediates the pair creation of black holes. Similarly, the wave function which describes the nucleation of a $dS$ de Sitter space from nothing is:

$$\psi_{dS} \propto \exp(-I_{ds}) \tag{3}$$

where $I_{dS} = -\dfrac{3 \cdot \pi}{2 \cdot \Lambda}$ is the action of the $S^4$ gravitational instanton which according to Lemos 'mediates' this nucleation. So, then the nucleation probability of the $dS$ space from nothing and then the $dS$ space with a pair of black holes from nothing is given by $|\psi_{dS}|^2$ and $|\psi_{inst}|^2$ respectively. This then allows us to state then that if we take the ratio of these two probabilities that we obtain the pair creation rate of black holes in the $dS$ background as

$$\Gamma \cong \eta \cdot \exp(-2 \cdot I_{inst} + 2 I_{dS}) \qquad (4)$$

For the Bogomol'nyi inequality approach[6,7] we modify a de facto 1+1 dimensional problem in condensed matter physics to being one which is quasi one dimensional by making the following substitution, namely looking at the Lagrangian density $\varsigma$ to having a time independent behavior denoted by a sudden pop up of a S-S' pair via the substitution of the nucleation 'pop up' time by[6,7]

$$\int d\tau \cdot dx \cdot \varsigma \to t_P \cdot \int dx \cdot L \qquad (5)$$

where $t_P$ is the Planck's time interval. Then afterwards, we shall use the substitution of $\hbar \equiv c \equiv 1$ so we can write

$$\psi \propto c \cdot \exp\left(-\beta \cdot \int L\ dx\right) \qquad (6)$$

A

## PPENDIX IV B: REDUCING THE GIVEN WAVE FUNCTIONAL TO HAVING GAUSSIAN FUNCTIONAL BEHAVIOR

We wish to give an argument as to how we obtain [6,7]

$$\Psi_{i,f}\left[\phi(\mathbf{x})\right]_{\phi \equiv \phi_{ci,cf}} = c_{i,f} \cdot \exp\left\{-\int d\mathbf{x}\, \alpha \left[\phi_{Ci,f}(\mathbf{x}) - \phi_0(\mathbf{x})\right]^2\right\}, \tag{1}$$

In both cases, we find that the coefficient in front of the wave functional in Eq. (1) is normalized due to error function integration

This is due to

We also found that in order to have a Gaussian potential in our wavefunctionals that we needed to have in both interpretations

$$\frac{(\{\ \})}{2} \equiv \Delta E_{gap} \equiv V_E(\phi_F) - V_E(\phi_T) \tag{2}$$

where for the Bogomol'nyi interpretation of this problem we worked with potentials (generalization of the extended Sine-Gordon model potential) [6,7]

$$V_E \cong C_1 \cdot (\phi - \phi_0)^2 - 4 \cdot C_2 \cdot \phi \cdot \phi_0 \cdot (\phi - \phi_0)^2 + C_2 \cdot (\phi^2 - \phi_0^2)^2 \tag{3}$$

We had a Lagrangian[15] we modified to be (due to the Bogomil'nyi inequality)

$$L_E \geq |Q| + \frac{1}{2} \cdot (\phi_0 - \phi_C)^2 \cdot \{\ \} \tag{4}$$

with topological charge $|Q| \to 0$ and with the Gaussian coefficient found in such a manner as to leave us with wave functionals [1,3,10] we generalized for charge density transport .This same Eq. (1) was more or less assumed in the Gaussian wavefunctional ansatz interpretation while

$$\Psi_f[\phi(\mathbf{x})]\Big|_{\phi \equiv \phi_{Cf}} =$$
$$c_f \cdot \exp\left\{-\int d\mathbf{x}\, \alpha \left[\phi_{Cf}(\mathbf{x}) - \phi_0(\mathbf{x})\right]^2\right\} \to \qquad (5)$$
$$c_2 \cdot \exp\left(-\alpha_2 \cdot \int d\tilde{x}[\phi_T]^2\right) \cong \Psi_{final},$$

and

$$\Psi_i[\phi(\mathbf{x})]\Big|_{\phi \equiv \phi_{Ci}}$$
$$= c_i \cdot \exp\left\{-\alpha \int d\mathbf{x}[\phi_{ci}(\mathbf{x}) - \phi_0]^2\right\} \to \qquad (6)$$
$$c_1 \cdot \exp\left(-\alpha_1 \cdot \int d\tilde{x}[\phi_F]^2\right) \equiv \Psi_{initial},$$

## APPENDIX V: EXTRA DIMENSIONS AND THE BREAK DOWN OF THE SEMI - CLASSICAL APPROXIMATION.THIS IS AN ILLUSTRATION OF THIS CONCEPT, AND NOTHING MORE

Here, I used equation 4.7 of the main text. For the first potential system, if we set xb=1, xa= - 1, and b = 10. (a sharp slope) for the scalar field boundary we have.

$$\alpha := \frac{.373}{1} \qquad (1)$$

This assumes a Gaussian wave functional of

$$\psi(x) := \exp(-\alpha \cdot \phi(x)) \qquad (2)$$

As well as a power parameter of

$$\nu := 9 \qquad (3)$$

Also, we are using, initially, a phase evolution parameter of

$$\phi(x) := \pi \cdot [\tanh[b \cdot (x - xa)] - \tanh[b \cdot (xb - x)]] \tag{4}$$

The first potential system is re scaled as

$$V1(x) := \frac{1}{2} \cdot (1 - \cos(\phi(x))) - \frac{1}{200} \cdot (\phi(x) - \pi)^2 \tag{5}$$

In addition, the following is used as a rescaling of the inner product

$$c1 := \frac{1}{\displaystyle\int_{-30}^{30} (\exp(-\alpha \cdot \phi(x)))^2 \cdot \frac{\pi^3}{3} \cdot x^5 \, dx} \tag{6}$$

$$c2 := \int_{-30}^{30} (\exp(-\alpha \cdot \phi(x)))^2 \cdot \frac{\pi^3}{3} \cdot x^5 \cdot (V1(x))^\nu \cdot |c1| \, dx \tag{7}$$

$$c3 := \left[ \int_{-30}^{30} (\exp(-\alpha \cdot \phi(x)))^2 \cdot \frac{\pi^3}{3} \cdot x^5 \cdot V1(x) \cdot |c1| \, dx \right]^\nu \tag{8}$$

$$c3b := \frac{c2}{c3} \tag{9}$$

Here,

$$C3b = .999 \tag{9a}$$

For the 2$^{nd}$ potential system, if we assume a sharp slope, i.e. b1 = b = 10, and

$$V2(x) := \frac{1}{2} \cdot \frac{(\phi a(x))^2}{1 + .000001 \cdot (\phi a(x))^3} \tag{10}$$

If

$$\phi a(x) := \pi \cdot [\tanh[b1 \cdot (x - xa)] - \tanh[b1 \cdot (xb - x)]] \tag{11}$$

and a modification of the 'Gaussian width' to be

$$\alpha 1 := \frac{.373}{30} \tag{12}$$

We do specify a denominator, due to a normalization contribution we write as

$$c1a := \frac{1}{\displaystyle\int_{-30}^{30} (\exp(-\alpha 1 \cdot \phi a(x)))^2 \cdot \frac{\pi^3}{3} \cdot x^5 \, dx} \tag{13}$$

$$c4 := \int_{-30}^{30} (\exp(-\alpha 1 \cdot \phi a(x)))^2 \cdot \frac{\pi^3}{3} \cdot x^5 \cdot (V2(x))^\nu \cdot |c1a| \, dx \tag{14}$$

In addition:

$$c5 := \left[ \int_{-30}^{30} (\exp(-\alpha \cdot \phi a(x)))^2 \cdot \frac{\pi^3}{3} \cdot x^5 \cdot V2(x) \cdot |c1a| \, dx \right]^\nu \tag{15}$$

We then use a ratio of

$$c5b := \frac{c4}{c5} \tag{16}$$

Here, when one has the six dimensions, plus the thin wall approximation:

$$C5b = 2.926\text{E-}3 \tag{17}$$

When one has three dimensions, plus the thin wall approximation

$$c6 := \int_{-30}^{30} (\exp(-\alpha 1 \cdot \phi a(x)))^2 \cdot \frac{1}{.25} \pi \cdot x^2 \cdot (V2(x))^\nu \cdot |c1b| \, dx \tag{18}$$

$$c7 := \left[ \int_{-30}^{30} (\exp(-\alpha \cdot \phi(x)))^2 \cdot \frac{1}{.25} \pi \cdot x^2 \cdot V2(x) \cdot |c1b| \, dx \right]^\nu \tag{19}$$

$$c7b := \frac{c6}{c7} \tag{20}$$

This leads to

$$c7b = .019 \tag{21}$$

When one has the thin wall approximation removed, via b1 = 1.5, one does not see a difference in the ratios obtained.

For the 3$^{rd}$ potential system, which is intermediate between the 1$^{st}$ and 2$^{nd}$ potentials if the b1 = b = 10 value is used, one obtains for when we have six dimensions

$$\alpha 1 := \frac{.373}{6} \tag{22}$$

As well as

$$V2(x) := \frac{1}{2} \cdot \frac{(\phi a(x))^2}{1 + .5 \cdot (\phi a(x))^3} \tag{23}$$

(When we have six dimensions)

$$C5b = 0.024 \tag{24}$$

(When we have three dimensions)

$$C7b = .016 \tag{25}$$

So, then one has C5b = .024, and C7b = .016 in the thin wall approximation

When b1 = 3 (non thin wall approximation)

$$C5b = .027 \tag{26}$$

(**Six dimensions**)

$$C7b = .02 \tag{27}$$

(**Three dimensions**)

Summarizing, if

$$V1(x) := \frac{1}{2} \cdot (1 - \cos(\phi(x))) - \frac{1}{200} \cdot (\phi(x) - \pi)^2 \quad = V1 \tag{28}$$

$$V2(x) := \frac{1}{2} \cdot \frac{(\phi a(x))^2}{1 + .000001 \cdot (\phi a(x))^3} \quad = V3 \tag{29}$$

$$V2(x) := \frac{1}{2} \cdot \frac{(\phi a(x))^2}{1 + .5 \cdot (\phi a(x))^3} \quad = V2 \tag{30}$$

One finally obtains the following results, as summarized below

|  | b=b1 = 10 | b1 = 3 | b1 = 1 |
|---|---|---|---|
| V1 ( 6 dim) | C3b = .999 | No data | No data |
| V3 ( 6 dim) | C5b = 2.926E-3 | No data | C5b = same value |
| V3 ( 3 dim) | C7b = .019 | No data | C7b = same value |
| V2( 6 dim) | C5b = .027 | C5b = .024 | No data |
| V2 ( 3 dim) | C7b = .02 | C7b = .016 | No data |

# APPENDIX VI: DECAY RATES, AND COSMIC NUCLEATION, I.E. PRESENTING A NEW WAY TO OBTAIN INITIAL EVOLUTION OF THE HUBBLE PARAMETER AND A RATE EQUATION

Garriga [33], assuming a nearly flat De Sitter universe also came up with an expression for the number density of particles per unit length (time independent)

$$n \approx \frac{1}{2 \cdot \pi} \cdot \sqrt{M^2 + e \cdot \frac{E_0^2}{H^2}} \cdot \exp(-S_E) \tag{1}$$

where for our purposes we would set

$$M \leq M_P \to 1 \tag{2}$$

We prefer instead to use an estimation of a nucleation rate per Hubble volume per Hubble time[29]

$$\in(t) \equiv \lambda_0 / (H(t)^4) \approx 1 \tag{3}$$

to show the influence an evolving Hubble parameter would have, in early times, without the complexity of predicting the $S_E$ ( a Euclidian action integral) which would be in our example a D+1 dimensional space knocked down to being quasi 1 dimensional in 'character'. We assume, also, rescaling of Planckian length to be unity where $\hbar \equiv c \equiv G \equiv 1$.

This leads to, then

# APPENDIX VII: PREDICTING HOW A SCALE FACTOR EVOLVES IN THE BEGINNING OF INFLATIONARY COSMOLOGY

I wish now to look at how the scale factor, $a$, changes in time, in a manner we view which will enable us to delineate Hubble parameter variations in the first few moments after creation. In doing this, we can note the typical value[34] ( as given by Dodelson)

$$a(t) \cong a_B \cdot \exp(H_B(t - t_B))  \qquad (1)$$

with $a_B$, $H_B$ and $t_B$ being scale factor, Hubble parameter, and time values at the end of an inflationary period of expansion. Needless to say, in doing this, we are not obtaining values of what the scale factor and Hubble parameter could be at the onset of inflation, which is a situation we wish to remedy. So we set

$$\tilde{a}_0 \equiv a_B \exp(H_B(t_P - t_B)) \qquad (2)$$

and afterwards approximate the evolution of phase after time $t_P$ via use of

$$\phi \equiv \tilde{\phi}_0 - \frac{m}{\sqrt{12 \cdot \pi \cdot G}} \cdot t \cong \tilde{\phi}_i \cdot \left( \exp(-\tilde{a}_0 \cdot t / \alpha) \approx 1 - \tilde{a}_0 \cdot t / \alpha \right) \qquad (3)$$

If we assume that $\tilde{\phi}_0 \cong \tilde{\phi}_i$, and that the time factors are small, we can state

$$\frac{m}{\sqrt{12 \cdot \pi \cdot G}} \cong \frac{\tilde{a}_0}{\alpha} \qquad (4)$$

as an order of magnitude estimate for the initial value of our scale factor at the beginning of inflation. So being the case, we move then to obtain a value for the initial evolution of the Hubble parameter via use of conformal time, with an Einstein equation

$$\ddot{\phi} + 2 \cdot a \cdot H \cdot \dot{\phi} + a^2 \cdot m^2 (\phi - \phi^*) = 0 \qquad (5a)$$

where the conformal time we write as

$$\tilde{t} \cong -\frac{1}{a(t) \cdot H} \qquad (5b)$$

which may be re written using ordinary time as

$$\ddot{\phi} + 3 \cdot H \cdot \dot{\phi} + m^2 (\phi - \phi^*) = 0 \qquad (5c)$$

which would lead to $\in (t) \equiv \lambda_0 / (H(t)^4) \approx 1$ [29] implying a nucleation rate evolution along the lines of

$$H \cong \frac{1}{3} \cdot \left(\frac{\tilde{a}_0}{\alpha}\right)^{-1} + \left(\frac{\tilde{a}_0}{\alpha}\right) \frac{m^2}{3} \cdot \left(1 - \frac{\phi^*}{\phi_0} \cdot \exp(\frac{\tilde{a}_0}{\alpha} \cdot \tilde{t})\right) \qquad (6)$$

implying

$$\lambda_0 (t_P + \delta \cdot t) \approx H^4 (t_P + \delta \cdot t) \qquad (7)$$

which for small times just past the initial value of $t \equiv t_P + \delta \cdot t$ leads to a nearly stable but increasing rate of the Hubble parameter right after a nucleation of a universe. This also leads to a phase change in behavior which I claim is motivated by the pre Planck time value of the Hubble parameter being set by (for times $t \leq t_P$)

$$H^2 \equiv \frac{8 \cdot \pi}{3} \cdot G \cdot V(\phi) \rightarrow \frac{8 \cdot \pi}{3} \cdot V(\phi) \qquad (8)$$

with the potential given by a washboard potential with a small driving potential proportional to $(\phi - \phi^*)^2$ which blends into Guths chaotic inflationary model for times $t_P + \delta \cdot t$.

# BIBLOGRAPHY